\documentclass[twocolumn,preprintnumbers,amsmath,amssymb]{revtex4-1}
\usepackage{txfonts}
\usepackage{amsmath}
\usepackage{newlfont}
\usepackage{amssymb}
\usepackage{multirow}
\usepackage{cases}
\usepackage[none]{hyphenat}
\usepackage{graphicx}
\usepackage{bm}
\usepackage{braket} 
\usepackage{epstopdf}
\usepackage{hyperref}
\def\beq{\begin{equation}}
\def\eeq{\end{equation}}
\def\beqa{\begin{eqnarray}}
\def\eeqa{\end{eqnarray}}
\newcommand{\eq}[1]{Eq. (#1)}

\newcommand{\fig}[1]{Fig. #1}

\newcommand{\pauj}[2]{ {{\sigma}^{#1}_#2}}

\newcommand{\ve}[1]{ \mathbf{\hat{r}}}

\newcommand{\ba}[1]{ \begin{equation} #1 \end{equation}}
\newcommand{\bag}[1]{ \begin{align} #1 \end{align}}

\begin{document}

\title{Parity Measurement in Ultrastrong Coupling Regime}
\author{Jing Yan~Haw$^1$, Yimin~Wang$^1$, and Valerio~Scarani$^{1,2}$}
\affiliation{$^1$Centre for Quantum Technologies, National University of Singapore, Singapore\\
$^2$Department of Physics, National University of Singapore, Singapore}

\date{\today}

\begin{abstract}
The measurement of the parity of two qubits is a primitive of quantum computing that allows creating deterministic entanglement. In the field of circuit quantum electrodynamics, a scheme to achieve parity measurement of two superconducting qubits has been proposed and analyzed under the usual rotating-wave approximation (RWA). We show that the same scheme can be carried over beyond this approximation, to the regime of ultrastrong coupling, with an improvement in the fidelity.
\end{abstract}

\maketitle
\section{Introduction}
The study of superconducting qubits coupled to a microwave resonator is known as circuit quantum electrodynamics (cQED) \cite{blais2004cavity,PhysRevA.75.032329}. Experiments have demonstrated the coherent control \cite{PhysRevLett.102.119901}, entanglement generation \cite{PhysRevA.81.062325} and readout \cite{PhysRevA.80.043840, PhysRevLett.102.200402} of qubits with high fidelity, making circuit QED one of the most promising architectures to achieve fault tolerant quantum computing \cite{nielsen2002quantum}.

One of the latest experimental achievements is the demonstration of \textit{ultrastrong coupling} \cite{forn2010observation,niemczyk2010circuit}, where the coupling strength between the qubit and the resonator is of the order of their resonance frequencies ($g/\omega_r\gtrsim0.1$). In this regime, physics is no longer captured by the usual rotating-wave approximation (RWA) and novel effects have been predicted \cite{PhysRevA.81.062131,PhysRevLett.105.263603}. In the context of quantum information processing, stronger coupling suggests faster dynamics, which is a very welcome development, since decoherence is one of the limiting factors in the performances of circuit QED. However, the translation is not so straightforward: for instance, two-qubit gates must be re-designed \cite{PhysRevLett.108.120501} because the usual architectures would suffer a rapid degradation of fidelity in this new regime \cite{ymwang}.

In this paper, we explore the extension to the ultrastrong coupling regime of another basic procedure: \textit{parity measurement}, which creates deterministic entanglement by extracting the overall parity of a two- or multi-qubit state. The existing proposals for parity measurement \cite{PhysRevA.81.040301,PhysRevA.82.012329,kockum2012undoing} are based on the \textit{dispersive qubit readout}: the qubits are strongly detuned from the cavity, the resonance frequency of the cavity is shifted depending on  the state of the qubits, and the joint information on single \cite{PhysRevA.77.012112} or multi-qubit \cite{PhysRevA.81.040301, bishop2009proposal} properties is obtained by homodyne measurement on a probe field transmitted through the resonator. This dispersive readout remains possible in the ultrastrong coupling regime \cite{zueco2009qubit}. As a result, we find that the proposed schemes for parity measurement reach even higher fidelity when $g/\omega_r$ increases.

The paper is organized as follow. In Sec. \ref{parity} we first present the drive-Rabi model in the dispersive regime. Next, we extend the two-qubit parity measurement to the ultrastrong coupling domain. In Sec. \ref{engen} by using the example of entanglement generation we discuss the deviations between the RWA and exact solution in different regimes of the parameters. Sec. \ref{conclu} summarizes our results.
\section{Parity Measurement}
\label{parity}
\begin{figure}[b]
\includegraphics[width=0.48\textwidth]{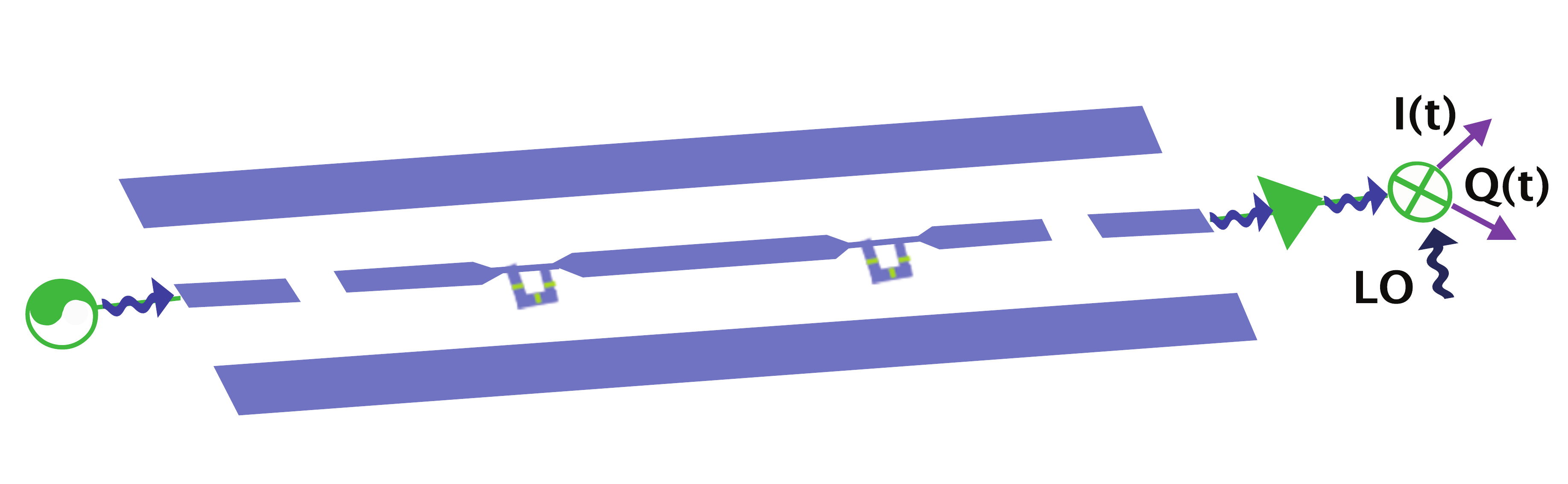}
\caption{(Color online) Schematic of the cQED setup. The microwave driving field is generated and transmitted through the resonator which is coupled ultrastrongly with the qubits. In the homodyne measurement, the microwave drive is amplified and mixed with a local oscillator (LO), where the in-phase $(I)$ and the quadrature $(Q)$ components of the transmitted field are measured.}
\label{schematic}
\end{figure}

In the dispersive regime with multiple qubits, the oscillator frequency exhibits a shift depending on the collective states of the qubits. This feature allows the joint readout of the multi-qubit state parity. By restricting to the two-qubit scenario, the objective of a two-qubit parity measurement is to distinguish the states between two orthogonal parity subspaces: $\cal{H}_+=$ span($\ket{gg},\ket{ee}$) and $\cal{H}_-=$ span($\ket{ge},\ket{eg}$). A perfect parity measurement projects the state of the system onto either subspace, without gaining any single-qubit information in order to preserve the superpositions in the post-measurement states. We consider a cQED system consisting of two superconducting qubits coupled to a transmission line resonator with a driving field applied to the input port (\fig{\ref{schematic}}). The corresponding Hamiltonian reads ($\hbar=1$) \cite{PhysRevLett.103.083601}
\beq
H=\omega_ra^{\dagger}a+\sum^2_{j}\frac{\omega_{aj}}{2}\pauj{z}{j}+g_j\pauj{x}{j}(a+a^\dagger)+\epsilon_m (a e^{i\omega_m t}+a^\dagger e^{ -i\omega_m t}).
\eeq
The first three terms are the usual Rabi model for atom-light interaction where $\omega_{aj}$ is the frequency of qubit $j$, $\omega_r$ is the frequency of the resonator and $g_j$ is the coupling strength between $j$th qubit and the resonator. The last term describes the feeding of the resonator by the classical field of amplitude $\epsilon_m$ at frequency $\omega_m$. \footnote{As we are operating with weak driving field throughout this paper, we have dropped the rapidly oscillating components in this last term.}

This drive-Rabi Hamiltonian, despite its simple form, is analytically non-trivial. Although the analytical solution of this Rabi Model has recently been found by Braak \cite{PhysRevLett.107.100401}, the general formula is in term of power series of a transcendental function. Since we are operating in the dispersive regime where the resonator and qubits frequencies are far detuned from each other compared to the coupling strength $g_j$, i.e.\ $g_j/\Delta_j\ll1$ and $g_j/\Sigma_j\ll1$, where $\Delta_j=\omega_r-\omega_{aj}$ and $\Sigma_j=\omega_r+\omega_{aj}$, a good approximate solution can be obtained by using the transformation \cite{zueco2009qubit,beaudoin2011dissipation,grptheoryrwa}
\ba{
U=\exp\left[\sum^2_j\frac{g_j}{\Delta_j} X^-_j+\frac{g_j}{\Sigma_j} Y^-_j\right],}
where $X^-_j=a\pauj{+}{j}- a^\dagger\pauj{-}{j}$, $Y^-_j=a\pauj{-}{j}- a^\dagger\pauj{+}{j}$. In the frame rotating at the drive frequency $\omega_m$, up to second order in the small parameters $g_j/\Delta_j$ and $g_j/\Sigma_j$, this transformation yields
\bag{
\label{effm}
H_\textrm{eff}&=\Delta_r a^\dagger a +\sum^2_j\frac{{\widetilde{\omega}_{aj}}}{2}\pauj{z}{j}
+\chi_j\pauj{z}{j}a^\dagger a\nonumber\\ 
&+\frac{\chi_j\pauj{z}{j}}{2}(a^2 e^{-i2\omega_m t}+a^{\dagger 2} e^ {i2\omega_m t})+ \epsilon_m (a+a^\dagger)\nonumber\\ 
&+\frac{g_j\epsilon_m}{\Delta_j}(\pauj{+}{j}+\pauj{-}{j})+\frac{g_j\epsilon_m}{\Sigma_j}(\pauj{+}{j}e^{i2\omega_m t} +\pauj{-}{j}e^{-i2\omega_m t})+J\pauj{x}{1}\pauj{x}{2},}
where $\Delta_r=\omega_r-\omega_m$, $\widetilde{\omega}_{aj}=\omega_{aj}+\chi_j$ is the Lamb-shifted qubit frequency, $\chi_j=g_j^{2}(1/\Delta_j+1/\Sigma_j)$ is the qubit state-dependent frequency shift and $J=g_1 g_2(1/\Delta_j-1/\Sigma_j)$ is the interqubit coupling mediated by virtual excitation of the field. The terms that oscillate with $\exp(\pm i2\omega_m)$ originate from the counter rotating terms $a\pauj{-}{j}$, $a^\dagger\pauj{+}{j}$ in the Rabi model. With the choice of $\omega_m\approx\omega_r$ for measurement on the qubits, we can safely ignore the qubit driving terms with amplitude $g_j\epsilon_m/\Delta_j$ and $g_j\epsilon_m/\Sigma_j$ \cite{PhysRevA.81.040301}. In order to focus on the entanglement generated by measurement, we drop the term proportional to $J$ since the possible measurement outcomes are the eigenstates of the interaction term $\pauj{x}{1}\pauj{x}{2}$. These measurement outcomes, on the contrary are not the eigenstates of the RWA's qubit-qubit interaction, which is proportional to $\pauj{+}{1}\pauj{-}{2}+ \mathrm{h.c.}$ \cite{PhysRevA.82.012329}.

In the Born-Markov approximation \cite{walls2008quantum}, the dynamics of the system in the presence of dissipation and dephasing is described by a Lindblad form master equation \cite{PhysRevA.74.042318}
\beq
\dot{\rho}=-\frac{i}{\hbar}[H_\textrm{eff},\rho]+\kappa\cal D[a]\rho+\gamma_{1j}\cal D[\pauj{-}{j}]\rho+\frac{\gamma_{\phi j}}{2}\cal D[\pauj{z}{j}]\rho,
\eeq
where $\cal D[L]\rho=(2L\rho L^\dagger - L^\dagger L \rho -\rho L^\dagger L)/2$. $\rho(t)$ is the density matrix of the total system, subjected to the loss of photons at rate $\kappa$, energy relaxation of the qubit $j$ at rate $\gamma_{1j}$ and the dephasing of the qubit $j$ at rate $\gamma_{\phi j}$. Following \cite{PhysRevA.74.042318} and neglecting the energy loss due to $\gamma_{1j}$, the intra-resonator field will evolve into a qubit-state dependent coherent states $\ket{\alpha_{xy}}$, $x,y=\{g,e\}$ with amplitude that satisfies
\beq
\label{parityf}
\dot{\alpha}_{xy}(t)=-i\epsilon_m-i[\Delta_r\alpha_{xy}(t)+\chi_{xy}(\alpha_{xy}(t)+\alpha^*_{xy}(t)e^{i2\omega_m t})]-\frac{\kappa}{2}\alpha_{xy}(t).
\eeq
Here $\chi_{xy}=\bra{xy}\chi_1\pauj{z}{1}+\chi_2\pauj{z}{2}\ket{xy}$. These qubit-state dependent amplitudes will act as the pointers in the measurement where the information of the states of the qubits can be inferred from. 

In the situation where $g_j/\omega_r\ll1$ and  $\Delta_j\ll\Sigma_j$, RWA is applicable to \eq{\ref{parityf}}. In this regime, we can approximate $\chi_j=g_j^{2}(1/\Delta_j+1/\Sigma_j)$ by $\chi^\textrm{RWA}_{j}=g_j^{2}/\Delta_j$ and average out the fast oscillating exponential term of frequency $2\omega_m\approx 2\omega_r$. This gives rise to the form
\beq
\dot{\alpha}^{\mathrm{RWA}}_{xy}(t)=-i\epsilon_m-i(\Delta_r+\chi^\textrm{RWA}_{xy})\alpha^{\mathrm{RWA}}_{xy}(t)-\frac{\kappa}{2}\alpha^{\mathrm{RWA}}_{xy}(t),
\label{parityrwa}
\eeq
where $\chi^\textrm{RWA}_{xy}=\bra{xy}\chi^\textrm{RWA}_{1}\pauj{z}{1}+\chi^\textrm{RWA}_2\pauj{z}{2}\ket{xy}$. This result in the limit of RWA agrees with the expressions in Refs. \cite{PhysRevA.74.042318,PhysRevA.81.040301}. Comparing these two differential equations for the coherent states amplitude of the exact case [\eq{\ref{parityf}}] and of the RWA case [\eq{\ref{parityrwa}}], we see that the former, apart from having an extra qubit state-dependent frequency shift of amplitude $g_j^2/\Sigma_j$, possesses an extra term involving the complex conjugate of the coherent state amplitude $\alpha^*$ modulated by an exponential term with frequency $2\omega_m$. This term originates from the two-photon process term in \eq{\ref{effm}}.
	
To perform a parity measurement, the measurement must be able to distinguish the parity subspaces without destroying the quantum superpositions within the subspace. By choosing identical qubits ($\omega_{aj}=\omega_a$, $j=1,2$) such that $\Delta_j=\Delta$, $\Sigma_j=\Sigma$, together with $g_j=g$ ($\chi_j=\chi$) and $\Delta_r=0$, we have $\chi_{xy}=\pm 2\chi$ for the states belonging to the even parity subspace $\cal{H}_+$ and $\chi_{xy}=0$ for the states belonging to the odd parity subspace $\cal{H}_-$ \footnote{Notice that the choice of $\Delta_1=-\Delta_2$ is invalid since this would lead to distinguishability of the states as $\Sigma_1\neq-\Sigma_2$ for this choice.}.
\begin{figure}[b]
\centering
 \includegraphics[scale=0.4]{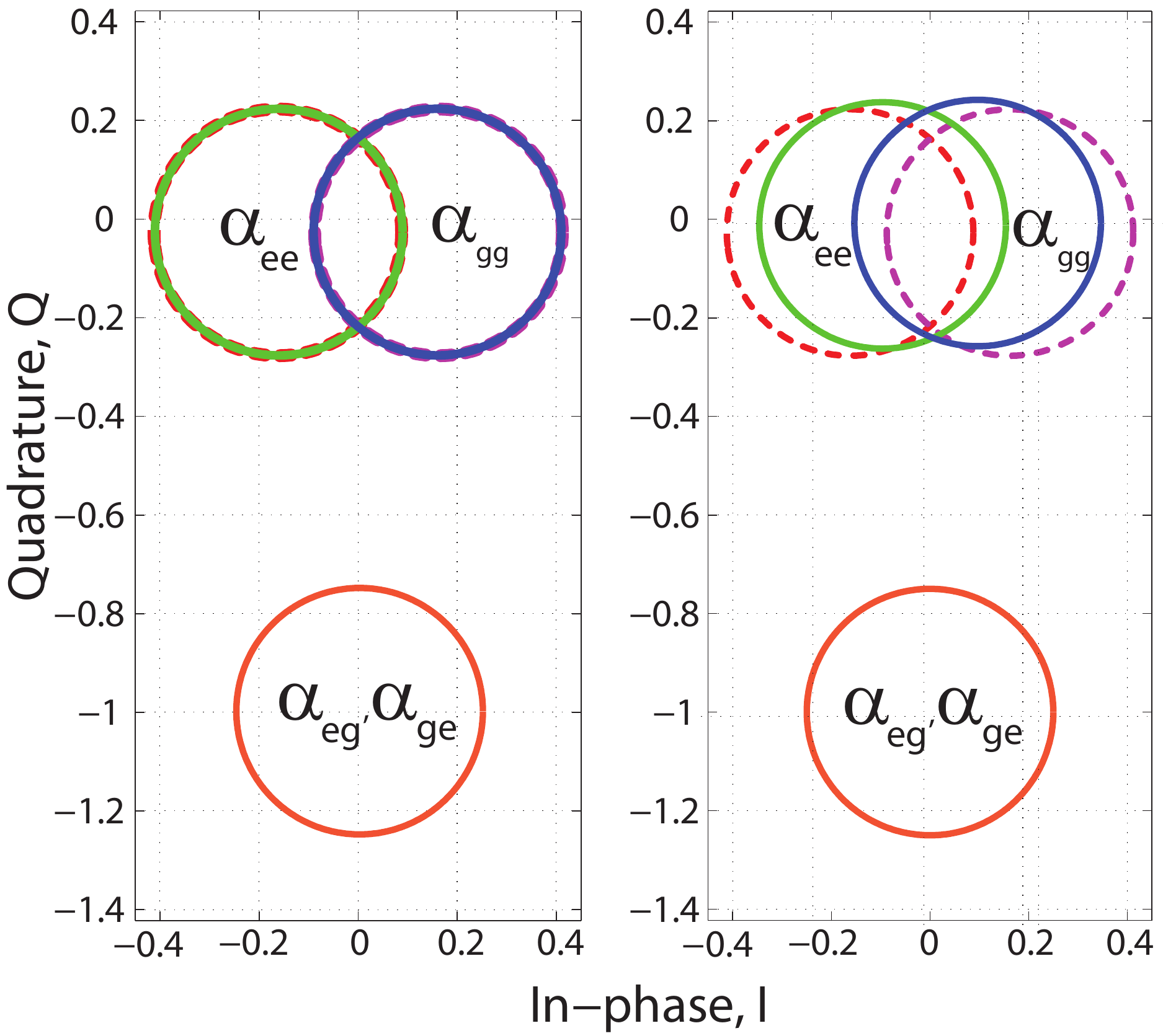} 
 	\caption{(Color online) The phase space illustration of the steady state solution of $\alpha_{xy}$ for both RWA (dashed lines) and exact (solid lines) with  $g/\omega_r=0.001$ (left) and $g/\omega_r=0.5$ (right). The parity of the qubits can be inferred by choosing the relative phase of the local oscillator to be $\phi=\pi/2$, which project the field state to the $Q$ quadrature. The parameters are $\Delta_r=0$, $\epsilon_m=0.5\kappa$ and $g=15\kappa$ with $\omega_r/2\pi=2$ GHz as fixed parameter.}
\label{fig:pcn}
	\end{figure}
The corresponding phase space of the steady states is illustrated in \fig{\ref{fig:pcn}}, where the quadrature $Q=\textrm{Im}[\alpha_{xy}(t)]$ is plotted against the in-phase component $I=\textrm{Re}[\alpha_{xy}(t)]$ of the transmitted field.

In the homodyne detection of the resonator field, with the proper choice of local oscillator phase $\phi$, one can measure the desired quadrature. The choice of $\phi=\pi/2$, which corresponds to the $Q$ quadrature, reveals only the information about the parity \cite{PhysRevA.81.040301, PhysRevA.82.012329}. However, there is still information in the quadrature orthogonal to the measurement, $I$ ($\phi=0$). As shown in \fig{\ref{fig:pcn}}, the overlay between the coherent states amplitudes $\alpha_{ee}$ and $\alpha_{gg}$ in even subspace $\cal{H}_+$ is not perfect. Hence this measurement is not optimal and leads to dephasing within the subspace $\cal{H}_+$. 

\begin{figure}[ht!]
\centering
\includegraphics[scale=0.24]{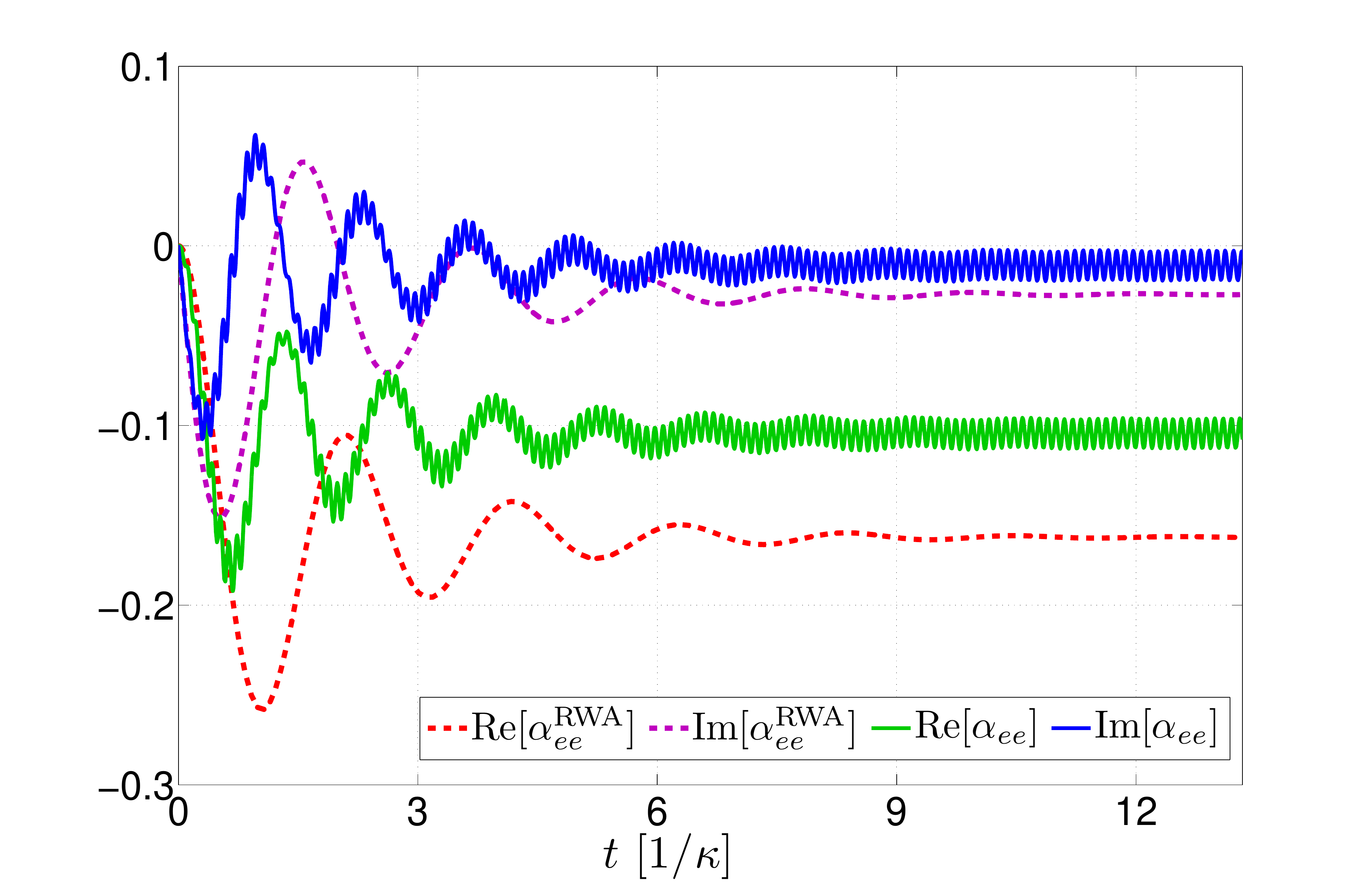}
\caption{(Color online) Dynamics of $\textrm{Re}[\alpha_{ee}(t)]$ and $\textrm{Im}[\alpha_{ee}(t)]$ vs time with coupling strength of $g/\omega_r=0.5$. The other parameters are the same as in \fig{\ref{fig:pcn}} (Notice that the color code is different from \fig{\ref{fig:pcn}}).}
\label{fig:ReImv}
\end{figure}

In \fig{\ref{fig:pcn}}, we compare the coherent states amplitudes in the regime of $g/\omega_r=0.001$ and $g/\omega_r=0.5$. Here we set $g/\kappa=15$, the same value considered in a recent study of feasibility of error correction benchmarks for parity measurement\cite{denhez2012quantum}. To keep the validity of the dispersive regime, we choose $g/\Delta=0.1$ \footnote{We also operate in the positive detuning regime $\Delta=\omega_{a}-\omega_r>0$ and assign $\omega_{a}$ as the tunable parameter to keep the ratio $g/\Delta$.} and set $\epsilon=0.5\kappa$ to stay much below the critical photon number $n_{\textrm{crit}}=(\Delta/2g)^2$ \cite{blais2004cavity}. When $g/\omega_r\ll 1$, the contributions of the counter rotating terms are insignificant, hence the exact solution and the RWA one are indistinguishable [left column of \fig{\ref{fig:pcn}}]. In the larger coupling strength regime where RWA is no longer applicable, the solutions differ from each other [right column of \fig{\ref{fig:pcn}}]. Deviations in both real and imaginary components of the coherent state amplitudes $\alpha_{ee}$ are depicted in \fig{\ref{fig:ReImv}}. In the exact case, the value of $\textrm{Re}[\alpha_{ee}(t)]$ is closer to zero. Since $\textrm{Re}[\alpha_{ee}(t)]=-\textrm{Re}[\alpha_{gg}(t)]$, this implies that the overlapping of the amplitudes increases. For the exact solution, there is also a slight increase in $\textrm{Im}[\alpha_{ee}(t)]$, which means that the separation between parity subspaces is somehow larger in the exact solution. Meanwhile, due to the modulation of the oscillating term in the exact solution \eq{\ref{parityf}}, small oscillations in the amplitudes persist even after the system reaches the steady state.	
\section{Entanglement Generation}
\label{engen}
To facilitate the understanding of these distinctions between RWA solution and exact solution, we focus on one specific task, namely the \textit{generation of entanglement from the separable states with parity measurement}. Consider the initial separable state $(\ket{g}+\ket{e})/\sqrt{2}\otimes(\ket{g}+\ket{e})/\sqrt{2}\otimes\ket{0}$, which corresponds to the first qubit, second qubit and vacuum resonator field respectively. Upon the displacement of the driving field, the joint atom-field state can be written as
\beq
\ket{\Psi}=\frac{1}{\sqrt{2}}\ket{\psi_+}\ket{\alpha_{eg}}+\frac{1}{2}[\ket{ee}\ket{\alpha_{ee}}+\ket{gg}\ket{\alpha_{gg}}].
\label{initial}
\eeq
The desired Bell states $\ket{\psi^+}=(\ket{eg}+\ket{ge})/\sqrt{2}$ and $\ket{\phi^+}=(\ket{ee}+\ket{gg})/\sqrt{2}$ can be obtained by $Q$ quadrature homodyne measurement on the transmitted field. To quantify the distinguishability between the entangled Bell states, the average fidelity of the measurement is defined as
\beq
F=P_\cal{H_+}F^{\phi_+}+P_\cal{H_-}F^{\psi_+},
\label{avef}
\eeq
where $P_\cal{H_+}$ and $P_\cal{H_-}$ are the success detection probabilities for the even and odd subspaces respectively. The expressions for the fidelity of the Bell states $F^{\phi_+}$ and $F^{\psi_+}$ are given in Appendix A.
\begin{figure}[b]   
\centering      
	 	 \includegraphics[scale=0.23]{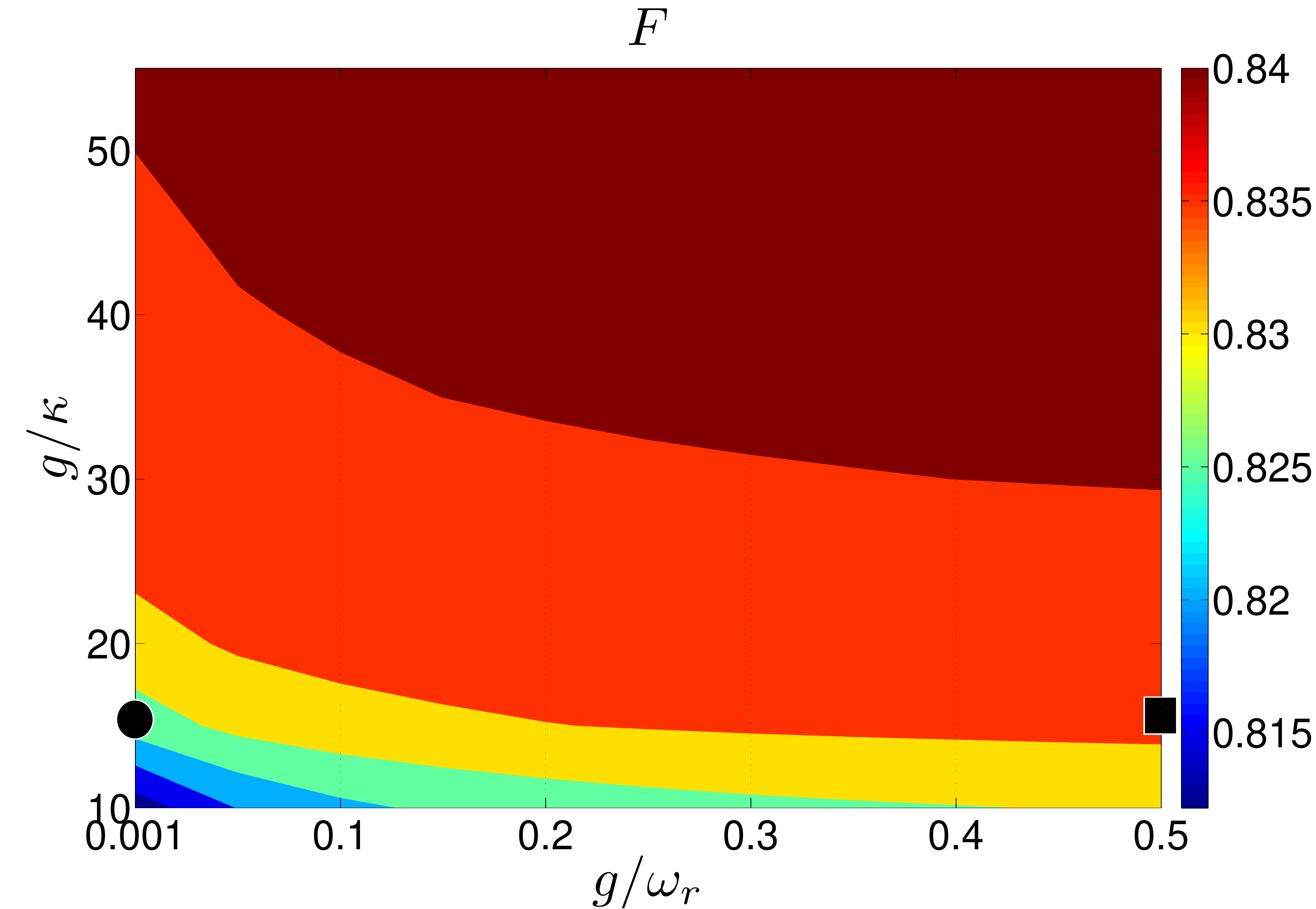}  
	 \includegraphics[scale=0.23]{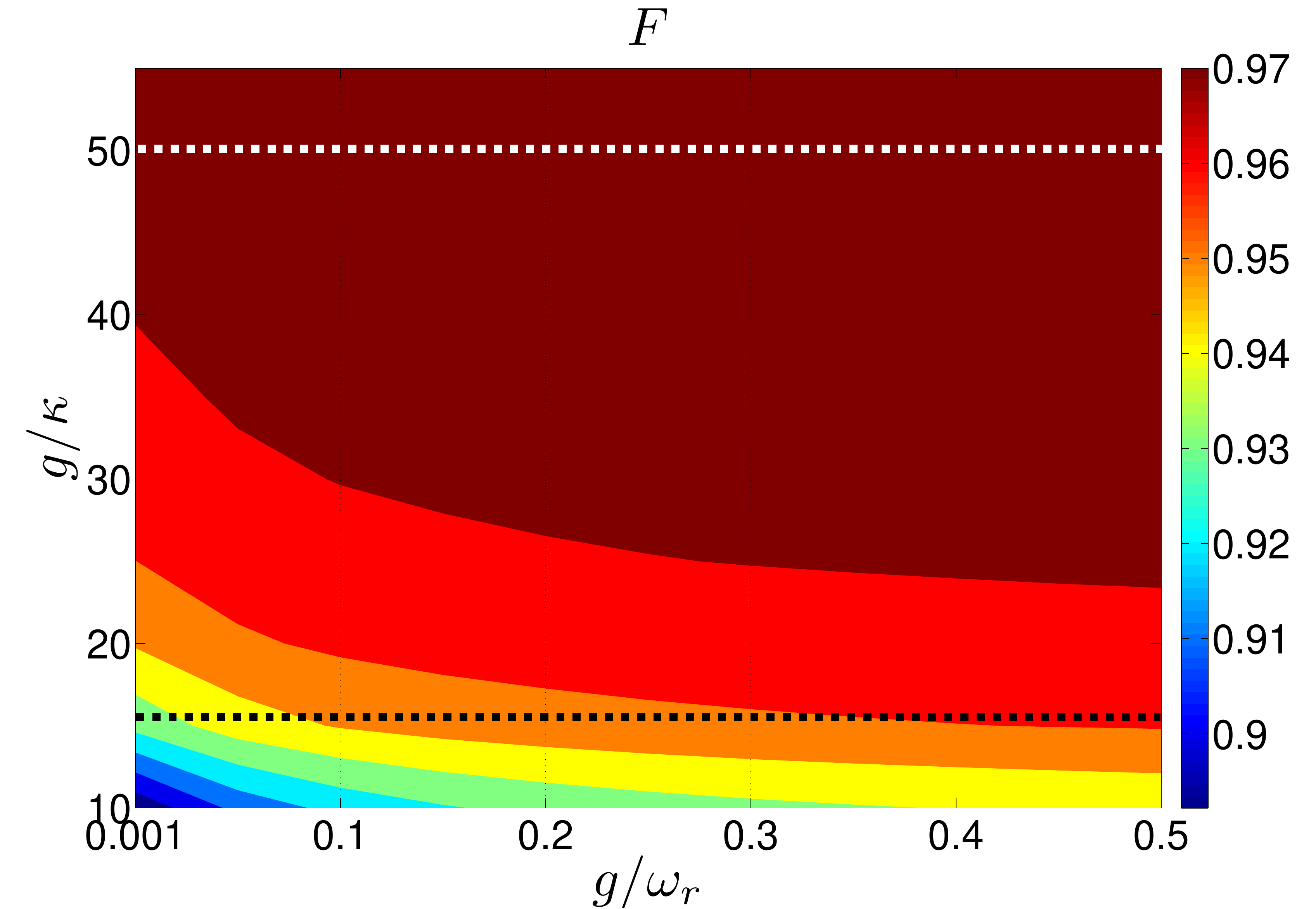}  
\caption{(Color online) The average fidelity $F$ of the parity measurement as a function of $g/\omega_r$ and $g/\kappa$ with $\epsilon=0.5\kappa$ (top) and $\epsilon=\kappa$ (bottom) where $\omega_r/2\pi=2$ GHz. The circle and square dots on the top figure ($g/\kappa=15$) correspond to phase space in \fig{\ref{fig:pcn}}.  The average fidelity of RWA ($F_\textrm{RWA}$) which corresponds to the value in the limit $g/\omega_r\rightarrow 0$ is not shown since it does not capture any dependence in $g/\omega_r$ and predicts a constant value.}
	\label{fig:FRWA}
\end{figure}
In \fig{\ref{fig:FRWA}} we plot the average fidelity of the entangled states as a function of both $g/\kappa$ and $g/\omega_r$, with $\omega_r/2\pi$ fixed as $2$ GHz. We observe that the average fidelity improves as the ratio $g/\kappa$ increases. This is because with higher ratio of $g/\kappa$, more information on the overall parity compare to the information on states within the subspaces is revealed \cite{PhysRevA.81.040301}. Notice that the maximum average fidelity for the choice $\epsilon=0.5\kappa$ is only about $0.84$. A higher maximum average fidelity of about $0.97$ can be achieved by choosing a larger driving field amplitude while remaining well below $n_{\textrm{crit}}$, i.e.\ $\epsilon=\kappa$. This is due to the fact that the coherent state amplitudes $\alpha_{eg}$ and $\alpha_{ge}$ of the odd parity subspace $H_\cal{-}$, which are proportional to the average number of photons \cite{PhysRevA.74.042318}, displace further away from the origin in the phase space as the photon number increases. Hence the separation between the parity subspaces increases, allowing the subspaces to be more distinguishable.

Let us now follow the physics of the problem when $g/\omega_r$ is varied. As the ratio $g/\omega_r$ further increases, as seen in the \fig{\ref{fig:FRWA}}, the average fidelity $F$ improves. This is due to the enhancement of the qubit state-dependent frequency shift $\chi$ in the exact case, where the factor $g/\Sigma=1/(\Delta/g+2\omega_r/g)$ in $\chi$ increases as $g/\omega_r$ rises. This implies that the qubits shift the resonator frequency by a larger amount, thus decrease the interaction between the qubits and the driving field at bare resonator frequency $\omega_r$. Hence the dephasing in the even parity subspace decreases \cite{PhysRevA.82.012329}.

\begin{figure}[b!]
	 \includegraphics[scale=0.26]{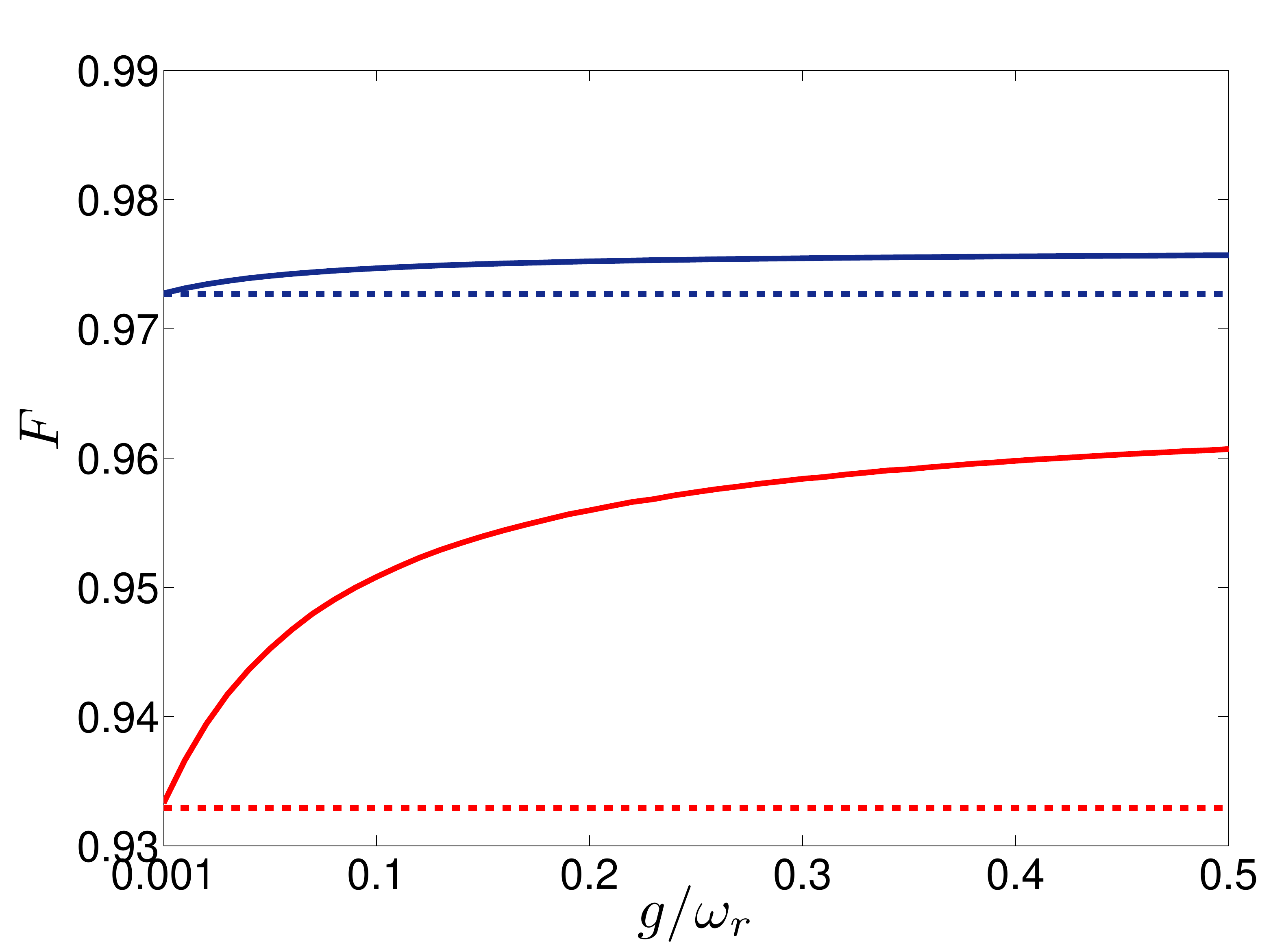}                
	\caption{(Color online) Average fidelity for RWA solution (dash) and exact solution (solid) with $g/\kappa=15$ (red) and $g/\kappa=50$ (blue), which correspond to the white and black dash lines in \fig{\ref{fig:FRWA}}.}
	\label{fig:FSL}
	\end{figure}
This enhancement in average fidelity is less significant for a higher ratio of $g/\kappa$. For comparison we plot the average fidelity for $g/\kappa=15$ and $g/\kappa=50$ respectively in \fig{\ref{fig:FSL}} (black and white dash line in the bottom column of \fig{\ref{fig:FRWA}}). The improvement in the average fidelity is significant for low ratio of $g/\kappa$ but not so much for a higher ratio of $g/\kappa$. This is basically due to the fact that at large value of $g/\kappa$, the overlapping between the states in the even subspaces is large and the parity measurement is close to optimal. As a result the advantage of the Rabi model over RWA in improving the average fidelity is less prominent, since the fidelity depends on the overlapping between the states in the subspaces and distinguisability between different parity subspaces.
\section{Conclusion}
\label{conclu}
In summary, we have explored the possibility of the dispersive parity measurement in the ultrastrong coupling regime in circuit QED. We have shown that, in general, the fidelity of the parity measurement is enhanced in this regime, due to the additional frequency shift that depends on the state of the qubits. 
\section{Acknowledgements}
We would like to thank Ji\v{r}\'{i}~Min\'{a}\v{r} and Peter H\"anggi for useful discussions. This work was supported by the National Research Foundation and the Ministry of Education, Singapore.
\appendix*
\section{Average Fidelity}
To derive the average fidelity for the parity measurement on the initial state, we follow the procedure outlined in Ref. \cite{PhysRevLett.96.240501}. Since we are measuring along the quadrature $Q$ ($\phi=\pi/2$), the conditional state of the initial state \eq{\ref{initial}} of the joint atom-field system for a measured $p$ value may now be expressed as
\beq
\ket{\Psi^C(p)}=\frac{C_{eg}(p)}{\sqrt{2}}\ket{\psi_+}+\frac{C_{ee}(p)}{2}\ket{ee}+\frac{C_{gg}(p)}{2}\ket{gg},
\label{cstate}
\eeq
where $C_{xy}(p)=G_{xy}(p)K_{xy}(p)$, with  $G_{xy}=(2/\pi)^{1/4}\exp[-(p-\textrm{Im}[\alpha_{xy}])^2]$ and $K_{xy}=\exp[-i\textrm{Re}[\alpha_{xy}](2p-\textrm{Im}[\alpha_{xy}])]$. After the homodyne detection of the resonator transmission, the joint system can be described by an unnormalized conditional density matrix $\rho^C(p)$ with the diagonal elements
\bag{
\bra{\psi_+}\rho^C(p)\ket{\psi_+}&=|C_{eg}(p)|^2/2,\nonumber\\
\bra{\psi_-}\rho^C(p)\ket{\psi_-}&=0,\nonumber\\
\bra{\phi_\pm}\rho^C(p)\ket{\phi_\pm}&=|C_{ee}(p)|^2/4\pm 1/8[C^2_{ee}(p)+C^{*2}_{gg}(p)].\nonumber\\
}
To distinguish between the odd and even parity states we define the midpoint of the subspaces
\beq
p_m=(\mathrm{Im}[\alpha_{ee}]+\mathrm{Im}[\alpha_{eg}])/2,
\eeq
and assign the post measurement results to $\cal{H_+}$ or $\cal{H_-}$ if $p>p_m$ or $p<p_m$. For $\cal{H_+}$, the success probability can be explicitly shown to be
\bag{
P_\cal{H_+}&=\int^\infty_{p_m} \textrm{d}p \textrm{Tr}[\rho^C(p)]\nonumber\\
&=\frac{1}{4}\left(\textrm{erfc}\left(\sqrt{2}(p_m-\textrm{Im}[\alpha_{eg}])\right)+\textrm{erfc}\left(\sqrt{2}(p_m-\textrm{Im}[\alpha_{ee}])\right)\right)\nonumber\\
&=\frac{1}{2},
}
similarly for $P_\cal{H_-}$, which is expected due to the symmetry between the subspaces over the quadrature $Q$. With this success probability, the fidelity of obtaining Bell state $\ket{\phi_+}$ from the $\cal{H_+}$ subspace can be calculated as
\bag{
F^{\phi_+}&=\frac{1}{P_\cal{H_+}}\int^\infty_{p_m}\textrm{d}p \bra{\phi_+}\rho^C(p)\ket{\phi_+}\nonumber\\
&=\frac{1}{4}\textrm{erfc}\left(\frac{\textrm{Im}[\alpha_{eg}]-\textrm{Im}[\alpha_{ee}]}{\sqrt{2}}\right)\nonumber\\
&+\frac{1}{8}e^{-2(b-ic)}\textrm{erfc}\left(\sqrt{2}\left(\frac{\textrm{Im}[\alpha_{eg}]-\textrm{Im}[\alpha_{ee}]}{2}-ib\right)\right)\nonumber\\
&+\frac{1}{8}e^{-2b(b+ic)}\textrm{erfc}\left(\sqrt{2}\left(\frac{\textrm{Im}[\alpha_{eg}]-\textrm{Im}[\alpha_{ee}]}{2}+ib\right)\right).
}\\
where $b=\textrm{Re}[\alpha_{ee}]$ and $c=\textrm{Im}[\alpha_{ee}]$. For the orthogonal Bell state of $\ket{\psi_+}$ from the $\cal{H_-}$ subspace, the fidelity is found to be
\bag{
F^{\psi_+}&=\frac{1}{P_\cal{H_-}}\int^{p_m}_{-\infty} \textrm{d}p \bra{\psi_+}\rho^C(p)\ket{\psi_+}\nonumber\\
&=\frac{1}{2}\textrm{erfc}\left(\frac{\textrm{Im}[\alpha_{eg}]-\textrm{Im}[\alpha_{ee}]}{\sqrt{2}}\right).
}
With these expressions, the average fidelity $F$ over the states $\ket{\psi_+}$ and $\ket{\phi_+}$ can be obtained by \eq{\ref{avef}}.
\bibliographystyle{prsty}
\bibliography{ultrastrong}

\end{document}